\documentclass[preprint,showpacs,preprintnumbers,amsmath,amssymb]{revtex4}

\usepackage{graphicx}
\usepackage{dcolumn}
\usepackage{bm}

\begin{document}

\preprint{APS/123-QED}

\title{Power thresholds of morphology dependent induced thermal scattering in
silica microresonators}

\author{Mikhail V. Jouravlev}
\email{jouravl@rambler.ru}
\affiliation{%
Raymond and Beverly Sackler Faculty of Exact Sciences, School of
Chemistry, Tel-Aviv University,
\\ Tel-Aviv, Ramat-Aviv, 69978, Israel}

\date{\today}

\begin{abstract}
Induced thermal scattering power thresholds have been calculated as
a function of size and laser pump frequency. The thermal coupling
coefficients of morphology dependent resonances were estimated by
asymptotic methods. The resulting power threshold is comparable with
experimental observations of thresholds of Raman lasing and thermal
instability in spherical silica resonators. Applications may include
the remote measurement of the temperature of aerosol droplets and
the stabilization of microcavity lasers against thermal oscillations
and temperature deviations on microcavity. A silica resonator can be
used as an IR sensor, as well as an additional tool for precisely
measuring the thermal conductivity and heat capacity of a target in
a microsphere by calculating of the thermal shifts of
eigenfrequencies in spectra of nonlinear scattering.
\end{abstract}

\pacs{42.60.Da, 42.55.Sa, 42.65.Sf}

\maketitle

Numerous stimulated nonlinear optical processes, including
Mandelstam-Brillouin and Raman scattering have been observed at very
low threshold laser pump intensities in liquid and solid
microresonators\cite{Lin1,Cantrell}. The main explanation for the
lowering of threshold conditions in microresonators is presented in
Ref.\cite{Kur2,Braun}. The nonlinear solid state silica
microresonators, having the "Whispering Gallery Modes" (WGM) with
high Q-factor $(Q=10^{8})$, act as low threshold microsphere Raman
lasers\cite{Vah1,Vah2}. The life time of photons of WGM is
significantly longer then in bulk solid. Due to the high photon
number in the effective volume of WGM and the long life time of
photons at "Morphology Dependent Resonance" (MDR) conditions, there
is a low Raman lasing threshold, which was observed in the
experiments with silica spheres having radius $R=40$ $\mu
m$\cite{Vah1}. The threshold power of the laser pump is $86$ $\mu W$
for a pump with wavelength $1.55$ $\mu m$, which is quite
low\cite{Vah1}. The absorbed pump power at MDR conditions changes
the eigenfrequency of a microresonator by heating the effective
volume of WGM. A periodic shift of the eigenfrequency provides
thermal instability in microresonators\cite{Gor1} and leads to a
bistability in the cavity\cite{Vah3,Vah4}. At the specific threshold
intensity of the laser pump and tuning conditions, the resonator
surface layer can significantly enhance the internal fields at MDR's
and efficiently provide optical thermal feedback for the internally
generated WGM, which leads to optical bistability and instability
effects\cite{Chang,Vah3,Vah4,Gor1}. The low threshold power of
thermal bistability and thermal oscillation in silica microresonator
have been reported for silica resonators with radii $100$ $\mu m$
and $150$ $\mu m$ at pump wavelength $0.63$ $\mu m$ \cite{Gor1}. The
high Q-factor of WGM provides the lowering threshold conditions for
the laser induced thermal scattering in the microresonator. Laser
stimulated thermal scattering in liquid balk have been reported in
Ref.\cite{Herm,Besp}. The goal of this article is to provide a
theoretical calculation of the threshold power of the laser induced
thermal scattering (LITS) in support of recent experimental efforts
in the area of thermal nonlinear effects in
microresonators\cite{Vah3,Vah4}. LITS is described by a system of
equations for nonlinear oscillators with partial wave
electromagnetic amplitudes $TE_{n}^{1}$ and thermal modes
$T_{n}^{0,1}$ in the resonator\cite{Bel1}. There are numerous
regimes of LITS depending on the combinations of the mode
overlapping and the amplitudes of partial electromagnetic and
thermal modes; these regimes include thermal instability, thermal
self-modulation and aperiodic thermal oscillation or
bistability\cite{Vah3,Vah4,Chang}. Experimentally, the effect of
thermal instability takes place in ring resonators\cite{Vah3,Vah4}
or fused silica spherical resonators\cite{Gor1}. There are two
interacting modes within the homogeneous thermal mode volume, one
being the pump mode and one being the resonant signal (anti-Stokes)
mode, i.e. the input and the output resonance conditions (MDR's).
The input resonance condition is satisfied for a broadband thermal
detuning $\eta$ of the pump input signal, which spans several high-Q
MDR's, whereas the output resonance condition is always satisfied,
since the bandwidth of LITS spans at least several high-Q MDR's. The
high-Q factor modes are modulated by the temperature oscillations of
the resonator due to the thermal shift of MDR's peaks near
$\omega_{f}$. The relevant Lorenzian is modulated by $\Omega_{T}$
and by the amplitudes corresponding to the thermal oscillations. In
the case of the interaction of two modes(one a thermal mode
$T_{1}^{1}$ and the second mode a WGM mode $TE_{n}^{1})$, the
threshold power is taken into account by applying the methods of
slowly varied amplitudes for the system of ordinary differential
equations describing the oscillations. The power threshold takes the
form \cite{Bel1}:
\begin{align}
P_{th}=\frac{\omega_{p}\rho C_{p} V}{2a_{\varepsilon}Q_{f}^{2}}
 \frac{1+\tau}{1\pm\gamma}\frac{1+\eta^{2}}{2\eta}
\end{align}
where:
\begin{align}
\gamma=\frac{3\pi\chi^{(3)}\varepsilon^{2}\rho
C_{p}}{\omega_{f}a_{\varepsilon}}
\end{align}
is a thermal and electromagnetic mode coupling coefficient and
\begin{align}
\tau=\frac{2KQ_{f}}{\rho
C_{p}\omega_{p}}\left(\frac{\mu_{j}}{R}\right)^{2}
\end{align}
The thermal anti-Stokes frequency has the form\cite{Bel1}:
\begin{align}
\Omega_{T}=\omega_{p} \left
[\frac{\xi_{f}^{2}/4+(\tau_{t}+\tau_{e})^{2}}
{1-\gamma}-\tau_{t}^{2} \right]^{1/2}
\end{align}
where: $\omega_{p}$ is the laser pump frequency,
$\eta=\xi_{f}/\xi_{0}$, $\xi_{f}=1-\omega_{f}^{2}/\omega_{p}^{2}$
and $\xi_{0}=2(\tau_{t}+\tau_{e})$ is the optimal detuning,
$\tau_{t}=K(\mu_{i}/R)^{2}/\rho C_{p}\omega_{p}$ and
$\tau_{e}=1/2Q_{f}$ are the dimensionless time of the thermal and
the electrical relaxation of the WGM, $V$ is the WGM's
volume\cite{Gor1}, $\mu_{i}$ are the roots of the boundary secular
equation for the spherical geometry of the resonator\cite{Bel1}. In
the result that follows, it is assumed that the pump field is
intense and undepleted, as opposed to the anti-Stokes field. We can
now derive the threshold incident power $P_{th}$ of the pump for
LITS using the basic relation for any Q-factor, namely, that a
Q-factor is the ratio of the field energy inside the mode to the
incident power, multiplied by the leakage rate. The best condition
for observation of LITS is provided by a spherical resonator from
fused quartz with a high Q-factor for WGM. The threshold incident
power $P_{th}$ of the pump for LITS can be derived using the basic
relation for any Q-factor namely,
\begin{equation}
\frac{1}{Q_{f}}=\left(\frac{1}{Q_{scat}}+\frac{1}{Q_{abs}}\right)_{f}
\end{equation}
where $(Q_{scat})_f$ and $(Q_{abs})_{f}$ are the Q-factors of
scattering loss and absorption loss in the f-mode. To be specific,
here we concentrate on the calculation of the threshold power of
LITS for modes whose amplitudes and resonance half-widths and
threshold of Raman lasing are given in \cite{Vah1,Cantrell}. The
calculated threshold power $P_{th}$ and the anti-Stokes frequency
$\Omega_{T}$ are presented by Fig.1-6, using the known values of
material parameters for the fused silica valid for the experiments:
the density of fused silica $\rho=2.21 [g/cm^3]$, the thermal
conductivity $ K=1.4\cdot10^{-2} [W/cmK]$, the specific heat
capacity $C_{p}=0.67 [Ws/gK]$, $a_{\varepsilon}=1.45\cdot10^{-5}
[K^{-1}]$,third-order susceptibility
$\chi^{(3)}=5\cdot10^{-15}[esu]$, $n=\varepsilon^{2}$, the
refraction index $n=1.46$. The threshold power of LITS is then
determined in two cases:(i) effective resonant heat absorption and
thermal mode effective overlapping, and (ii) heat exchange on the
surface of the resonator. The next assumption takes into account the
small losses of the electromagnetic mode, which corresponds to the
WGM's having a high-Q factor\cite{Gor1,Gor4}. Optimal tuning
conditions for observating of LITS can be created using a  spherical
resonator from fused silica with a high Q-factor of $Q_{f}=$
$10^{7}$ \cite{Gor4}. The Q-factor in Eq.(1) and Eq.(4) leads to the
expected linear dependence of the loss rate. The absorbed pump
energy depends on the laser pump power, while the Q-factor leads to
LITS through the contribution of stored energy within the
electromagnetic mode volume, which heats the system and raises the
entropy through the contribution of the temperature dependence of
the index of refraction on the scattering volume. This dependence is
reminiscent of stimulated thermal scattering in liquid bulk and is a
important factor in understanding the single-photon absorption rate
for a molecular system with a chemical impurity. Thus, the threshold
power of the incident laser pump is inversely proportional to the
thermal coupling of partial modes $\gamma$, and to the Q-factor of
the pump mode squared.
\begin{figure}[htb]
\centerline{\includegraphics[width=8cm]{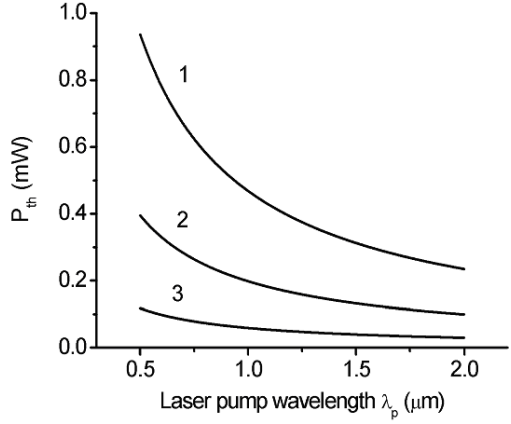}} \caption{The
threshold power of LITS depends on resonant wavelength of wide range
of gas laser and solid state semiconductor lasers. Radius of fused
silica microsphere: 1. $R=20$ $\mu m$, 2. $R=15$ $\mu m$, 3. $R=10$
$\mu m$.,$Q_{f}=10^{8}$, $TE_{n}^{1}-T_{1}^{1}$,$n\approx\rho$.}
\end{figure} The LITS power threshold evaluated for the optimal detuning
$\xi_{0}$ in MDR's as a function of resonant pump wavelength is
presented in Fig.1. The interacting electromagnetic modes
$E_{n}^{1}$ corresponding to WGM's $n \approx \rho$, where:
$\rho=2\pi R/\lambda_{p}$ is the size parameter,$R$- radius of the
resonator,$\lambda_{p}$ is the pump wavelength.
\begin{figure}[htb]
\centerline{\includegraphics[width=8cm]{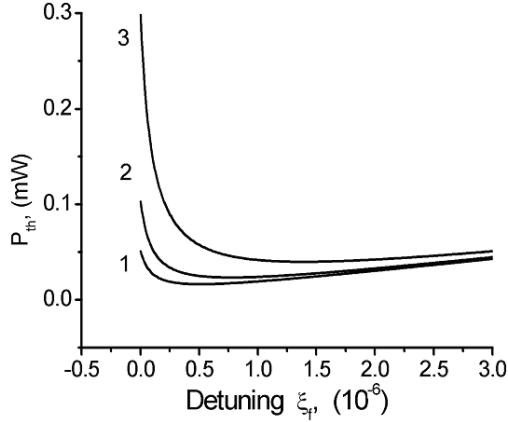}} \caption{The
threshold power of LITS as a function of detuning $\xi_{f}$ (radius
$R=35$ $\mu m$) exited the MDR's: 1.$ \lambda_{p}=0.532$ $\mu m$,
$TE_{600}^{1}-T_{1}^{0}$ modes, 2. $\lambda_{p}=0.840$ $\mu m$,
$TE_{300}^{1}-T_{1}^{0}$ modes. 3. $\lambda_{p}=1.55$ $\mu m$,
$TE_{210}^{1}-T_{1}^{0}$.}
\end{figure}
\begin{figure}[htb]
\centerline{\includegraphics[width=8cm]{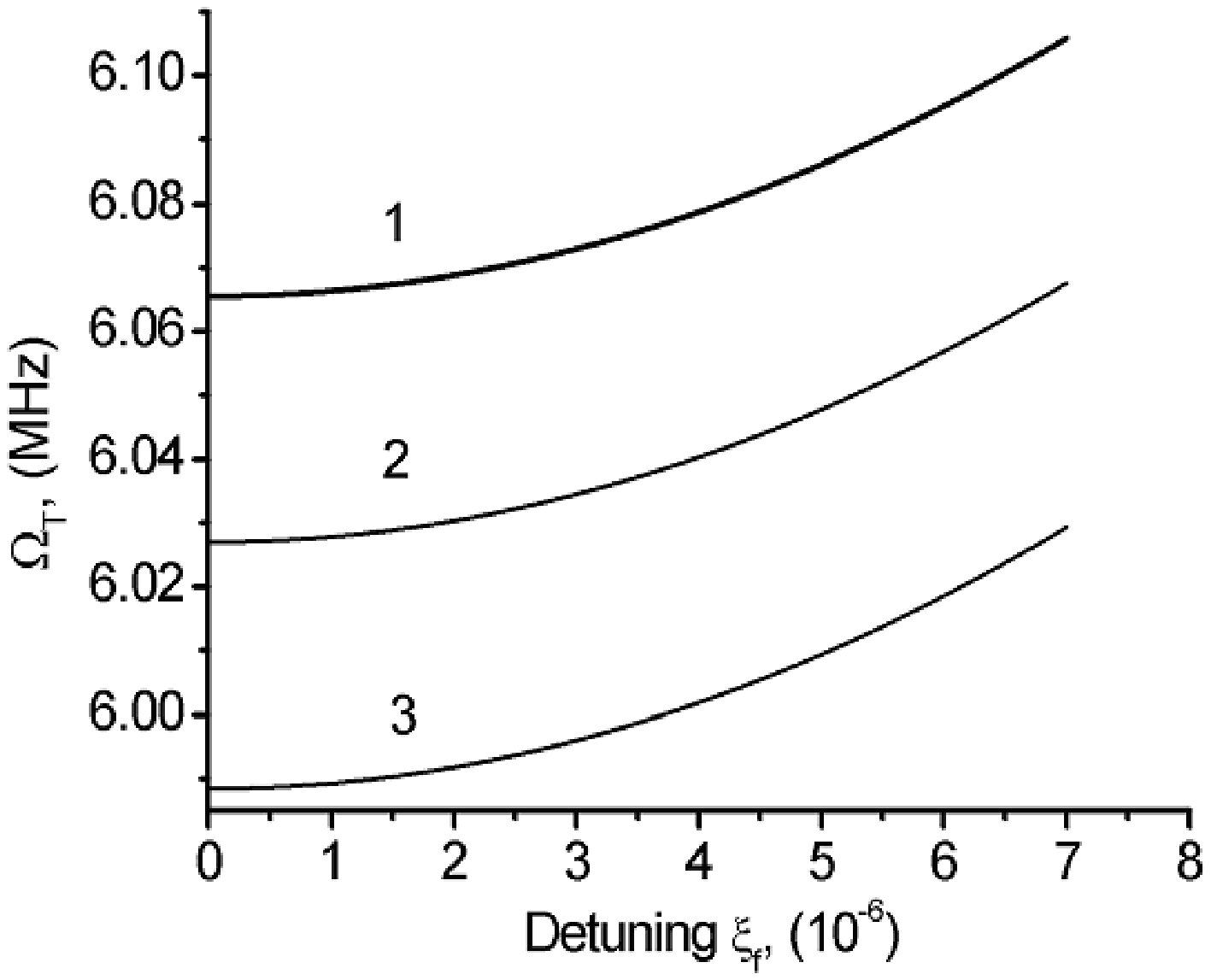}} \caption{The
thermal combination anti-Stokes frequency of LITS as a function of
detuning $\xi_{f}$ (radius $R=35$ $\mu m$) exited the MDR's: 1.$
\lambda_{p}=0.532$ $\mu m$, $TE_{600}^{1}-T_{1}^{0}$ modes, 2.
$\lambda_{p}=0.840$ $\mu m$, $TE_{300}^{1}-T_{1}^{0}$ modes. 3.
$\lambda_{p}=1.55$ $\mu m$, $TE_{210}^{1}-T_{1}^{0}$.}
\end{figure}
The threshold power and thermal combination anti-Stokes frequency
with dependence from detuning $\xi_{f}$ presented in Fig.2 and
Fig.3. As illustrated in Fig.2-3, the optimal tuning condition
provides the minimum of incident laser power for LITS inside a
microresonator. It corresponds to the detuning of the effective
two-mode interaction with $\xi_{f}$ comparable with the $\xi_{0}$.
The threshold power has a minimum at the optimal tuning of the pump
wavelength.
\begin{figure}[htb]
\centerline{\includegraphics[width=8cm]{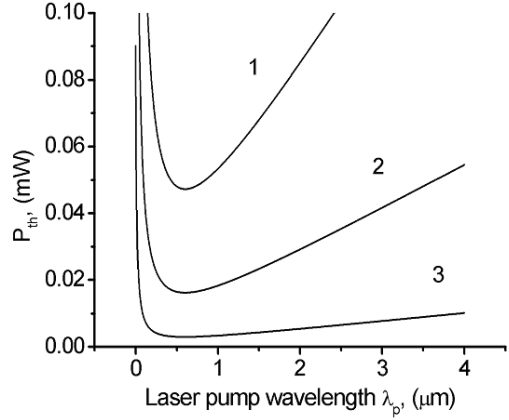}} \caption{ The
minimum of threshold of LITS for silica microsphere at the laser
pump $\lambda_{p}=0.532$ $\mu m$ with radii 1.$R=70$ $\mu m$, 2.
$R=45$ $\mu m$, 3.$R=50$ $\mu m$.}
\end{figure} The threshold power of LITS, as a function
that depends on the laser pump wavelength, is presented in Fig.4. It
was included in the detuning function $\xi_{f}$. The minimum of the
threshold intensity is achieved by MDR's tuning at
$\lambda_{p}=0.532$ $\mu m$ and consists of $20$ $\mu W$. It is
consistent with the experimental data\cite{Gor1} and less then that
of the threshold for Raman lasing\cite{Vah1}. The computed threshold
for input power in a microresonator varies from $20$ to $50$ $\mu W$
for detunings of 5 to 15 linewidths from a $TE$ resonant mode.
\begin{figure}[htb]
\centerline{\includegraphics[width=8cm]{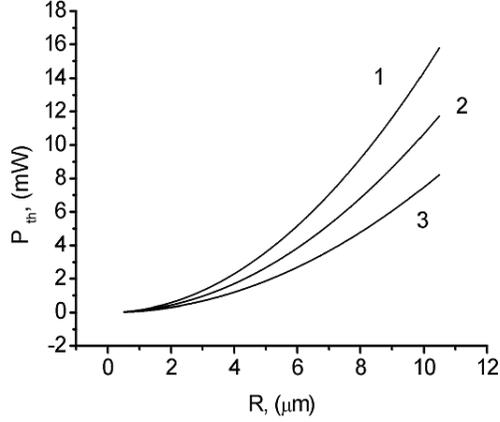}} \caption{The
threshold power of LITS for fused silica microsphere.
$Q_{f}=10^{7}$, $\lambda_{p}=1.550$ $\mu m$,1.
$TE_{230}^{1}-T_{1}^{0}$, 2. $TE_{170}^{1}-T_{1}^{0}$,
3.$TE_{120}^{1}-T_{1}^{0}$.}
\end{figure}
\begin{figure}[htb]
\centerline{\includegraphics[width=8cm]{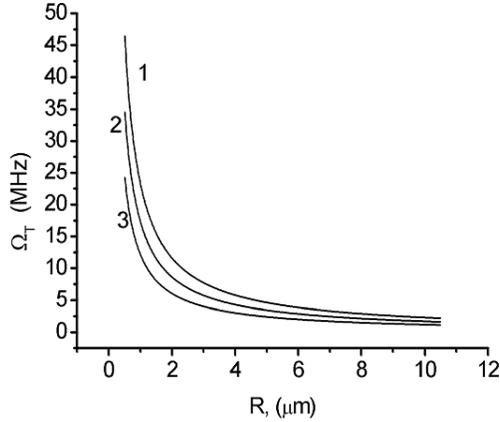}} \caption{The
thermal combination anti-Stokes frequency of LITS for fused silica
microsphere $Q_{f}=10^{7}$, $\lambda_{p}=1.550$ $\mu m$,
1.$TE_{230}^{1}-T_{1}^{0}$, 2.$TE_{170}^{1}-T_{1}^{0}$,
 3.$TE_{120}^{1}-T_{1}^{0}$.}
\end{figure}
Our new result for threshold power and $\Omega_{T}$ calculated by
the formulas (1)-(4) are presented in Fig. 5 and Fig. 6 allowing us
to compare the threshold and combination frequencies of relevant
experimental parameters for the small spherical silica particles
suspended or spraying in the atmosphere. The $\Omega_{T}$ for the
microspheres with radii $R = 2 \mu m \div 10  \mu m$ (the probe
particles in the atmosphere) is smaller
 than $5$ $MHz$ in the regime of time of pump pulse: $
\tau_{p}\gg \tau_{t}\gg \tau_{e}$. The thermal combination frequency
$\Omega_{p}$ for the silica spheres with radius of $5 \mu m$  has a
value between $5$ $MHz$ and $10$ $MHz$. For a micrometer-sized
silica resonator the thermal mechanisms become important for time
scales on the order of $\tau_{t}$ for thermal relaxation, a few
microseconds. It can occur in the Raman and Mandelstam-Brilluien
spectra in the silica spheres and can provide thermal modulation of
Raman and Mandelstam-Brilluien scattering amplitudes in the
experiments with lasing in silica spheres.
 The combination of anti-Stokes frequency of LITS can be varied in a wide range from
 a value of combination Rayleigh frequencies (Rayleigh thermal scattering\cite{Herm,Besp}) to striction
 Brillouin frequencies (Mandelshtam-Brillouin scattering\cite{Cantrell}).
 Such a wide variation of values of anti-Stokes frequency $\Omega_{T}$ is provided by the distributed
 nonlinear system, like a resonator. The calculated threshold power of LITS is significantly less than the
thresholds of Mandelstam-Brillouin scattering, which is less than
$160$ $W$\cite{Cantrell}. Thus, it is necessary to take into account
LITS on small particles in the scattering experiments. It is found
that the thresholds of parametric LITS in silica microspheres are
amenable to experimental observation, because this threshold is
commensurate with the threshold of stimulated Raman scattering on
microspheres or also with the threshold Raman lasing in the volume
of WGM at resonator\cite{Vah1}. Furthermore, we may assume that the
normalized gain of Raman lasing can be modulated by the excitation
of thermal mode of higher order by partial electromagnetic waves.
Thermal modes and the surface WGM mode pairs with strong nonlinear
coupling have low thresholds and thermal combination frequencies
$2.5\div50$ MHz. As was shown in\cite{Pell} there is a line
broadening of the modes if the microsphere is doped with latex
nanoparticles. The concentration of any absorbing or scattering
nanoparticles can be estimated by power thresholds and Anti-Stokes
frequency shifts with size parameters. It provides the experimental
tool for controlling the polymerization on the surface of
microdroplets and microresonators. The threshold for LITS will
decrease due to losses incurred by absorbing nanoparticles. However,
if the nanoparticles are present on the surface of resonators, the
effect of stimulated Raman scattering, Raman lasing and
line-broadening may keep the thermal modulation of Q-factors intact
and cause the splitting of resonant modes. In this case the integral
coefficients of interaction between Anti-Stokes and pump ( or signal
and idler) modes may actually increase. This intriguing possibility
leads to further lowering of the threshold conditions of LITS, whose
practical aim may be the creation of broad-band microsphere thermal
sensors. This work has provided the asymptotical an approach for
threshold conditions of LITS. In order to run more exact
calculations, it may be necessary to develop a detailed theory of
thermal and electromagnetic mode overlapping. The theory presented
above could be used for resonators of any geometrical form, but it
while necessary to provide the more complicated solution for
boundary problems of diffraction and thermal conductivity theory,
even though the basic physics of our solution does not change.
Threshold measurements can be an effective tool for chemical
substance detection and for the creation of IR sensors.

\end{document}